\documentclass[aps,prb,reprint,groupedaddress]{revtex4-2}

\usepackage{graphicx}
\usepackage{bm}

\begin{document}


\title{Staggered spin susceptibility at a two-dimensional antiferromagnetic quantum critical point}

\author{Yutaka Itoh}
\affiliation{Department of Physics, Graduate School of Science, Kyoto Sangyo University, Kamigamo-Motoyama, Kita-ku, Kyoto 603-8555, Japan}
 
\date{\today}

\begin{abstract}
We report on the finite temperature staggered spin susceptibility $\chi(Q)$ as a function of the mode-mode coupling constant $y_1$ in the self-consistent renormalization theory of two-dimensional antiferromagnetic spin fluctuations with zero-point quantum fluctuations just at the quantum critical point ($y_0$ = 0). 
We find that the value $y_1$ = 0.1 is a criterion to classify the effect of the zero-point spin fluctuations on the temperature dependence of $\chi(Q)$
into a Curie law for weak $y_1 < $ 0.1 and a Curie-Weiss type or a power law type for strong $y_1 > $ 0.1.
The absence of a Curie-Weiss temperature can serve as an identifying criterion for QCP ($y_0$ = 0) in systems with weak mode-mode coupling ($y_1 <$ 0.1). 
Experimental application on the $y_1$ classification is shown to several itinerant layered antiferromagnetic systems through an analysis of nuclear spin-lattice relaxation rates. 
\end{abstract}


\maketitle

\section{Introduction}
One of the key ingredients of itinerant magnetism is the mode-mode coupling constant $y_1$ of spin fluctuations~\cite{MT}.
Although it is possible to estimate $y_1$ for itinerant ferromagnets through the analysis of Arrott plots from high-field magnetization measurements~\cite{TY},
there is no standard experimental protocol for estimating $y_1$ in itinerant antiferromagnets.
The staggered spin susceptibility $\chi(Q)$ ($Q$ is an antiferromagnetic wave vector) can be measured by NMR and neutron scattering experiments.  
In experimental research, there has been a need to have practical theoretical expressions of the staggered spin susceptibility $\chi(Q)$ using $y_1$.

Two-dimensional antiferromagnetic spin fluctuations have attracted great attention in research on the mechanisms of high-temperature superconductivity~\cite{MU1,MU2,SCA,DAI}.  
Itinerant antiferromagnetism at a quantum critical point (QCP) in two spatial dimensions has been an active topic~\cite{QCP1,QCP2}.
A recent advancement in spin fluctuation theory is the recognition of the influence of zero-point quantum fluctuations at and near a QCP.  
The zero-point spin fluctuations play a significant role in the quantum critical phenomena of itinerant electron magnetism at a two-dimensional antiferromagnetic QCP~\cite{WM, IM1, TM}. 
The two-dimensional quantum critical behavior of the staggered spin susceptibility $\chi(Q)$ is logarithmic temperature dependence near absolute zero temperature~\cite{WM, IM1, TM, Millis}.  
The logarithmic divergence of $\chi(Q)$ holds only at extremely low temperatures~\cite{WM, IM1, TM, Millis}. 
The finite temperature $\chi(Q)$ at the QCP is of a Curie-Weiss type in the self-consistent renormalization (SCR) theory with zero-point quantum fluctuations~\cite{WM, IM1, TM}.
The logarithmic divergence near absolute zero and the finite temperature Curie-Weiss law are characteristic of $\chi(Q)$ at the two-dimensional antiferromagnetic QCP in the SCR theory with the zero-point fluctuations.
These are different from a Curie law regarding $\chi(Q)$ at the QCP in the previous SCR theory, which considers only thermal fluctuations~\cite{MTU}.

Although the numerical $\chi(Q)$ is called "Curie-Weiss behavior"~\cite{WM, IM1, TM}, 
it has been unclear how the Curie-Weiss law is parameterized in the SCR theory with the zero-point fluctuations.
The Curie-Weiss law of $\chi(Q)$ in nearly antiferromagnetic systems without the zero-point fluctuations~\cite{MTU} was expressed in terms of the SCR parameters~\cite{Itoh}. 
Phenomenological simple expressions of $\chi(Q)$ are convenient for experimental estimations of the spin fluctuation parameters. 
It is crucial for experimental studies whether $\chi(Q)$ is a Curie law or a Curie-Weiss law, because the absence of the finite Curie-Weiss temperature has served as identification of the QCP~\cite{Itoh,Itoh2,Ishida}.    

In this paper, according to the Watanabe-Miyake SCR theory with the zero-point spin fluctuations~\cite{WM}, we study the SCR equation of $\chi(Q)$ as a function of the mode-mode coupling constant $y_1$ at the two-dimensional antiferromagnetic QCP. 
We find that the effect of the zero-point spin fluctuations decreases with decreasing the value of $y_1$.
The temperature dependence of $\chi(Q)$ at finite temperatures depends on the value of $y_1$.   
$\chi(Q)$ changes from a Curie law for weak $y_1 < $ 0.1 to a Curie-Weiss type or a power law type for strong $y_1 > $ 0.1.
The $y_1$ classification is applied to actual itinerant layered compounds by an analysis of experimental nuclear spin-lattice relaxation rates. 

\section{SCR theory with zero-point quantum fluctuations at a two-dimensional QCP : two asymptotic solutions}

Energy units for temperature $T$ and frequency $\omega$ are set by $k_{\mathrm B}$ and ${\hbar}$ = 1. 
$\chi(Q)$ is in units of $(2\mu_\mathrm{B})^2$.
The notations conform with the Watanabe-Miyake SCR theory~\cite{WM}.  

The dynamical spin susceptibility $\chi(Q+q, \omega)$ peaked at $Q$ is expanded as $1/\chi(Q+q, \omega)$ = 2$\pi T_A\{(y + x^2) - i\omega/2\pi T_0\}$ 
where the reduced wave number $x$ = $q/q_B$ ($q_B$ is the effective Brillouin zone boundary), the reduced inverse staggered spin susceptibility $y$ = 1/2$T_A\chi(Q)$ = 1/$(\xi q_B)^2$ ($\xi$ is the antiferromagnetic correlation length), the frequency spread $T_0$ and the spatial spread $T_A$ of the spin fluctuation spectrum.  

For the given values of the mode-mode coupling constant $y_1$ and just at the QCP distance $y_0$ = 0, 
the Watanabe-Miyake SCR equation of $y$ against the reduced temperature $t$ = $T/T_0$ is
\begin{equation}
y = \frac{y_1}{2}\Big[ (y{\rm ln}y - c y) +\frac{t}{2}\Big\{{\rm ln}\frac{x_c^2 + y}{y}-{\rm ln}\frac{x_c^2 + y + t/6}{y + t/6}\Big\}\Big],
\label{eq1} 
\end{equation} 
where $x_c$ = $q_c/q_B$ is the cut-off wave number and $c$ = 1 + 2${\rm ln}x_c$~\cite{WM,IM1}. 
The logarithmic term of $(y{\rm ln}y - c y)$ due to the zero-point quantum fluctuations is characteristic of the two dimensional antiferromagnetic QCP. 
The term of $(t/2)\{\cdots\}$ results from the thermal fluctuations. 
 
In the low temperature limit of $t$ $\rightarrow$ 0, the SCR Equation~(\ref{eq1}) with finite $y_1$ leads to
\begin{equation}
y \sim \frac{-t{\rm ln}|{\rm ln}t|}{2{\rm ln}t},
\label{QCP} 
\end{equation}
which is derived from the cancellation of the zero-point fluctuations and the thermal fluctuations as $y\sim$ 0 in Eq.~(\ref{eq1})~\cite{WM,IM1}.
An important feature of Equation~(\ref{QCP}) is that it does not depend on $y_1$. 
$\chi(Q)$ exhibits logarithmic divergence near absolute zero for all finite values of $y_1$. 

At finite temperatures of $t$ = 0.001--0.30, in the weak coupling limit of $y_1$ $\rightarrow$ 0, then $y \ll t$, the SCR Eq.~(\ref{eq1}) leads to 
\begin{equation}
y \sim \frac{-y_1 {\rm ln}y_1}{4}t
\label{CL} 
\end{equation}
and then
\begin{equation}
\chi(Q) \sim \frac{2}{-y_1 {\rm ln}y_1}\frac{T_0}{T_AT}.
\label{XQ} 
\end{equation} 
This is a Curie law of $\chi(Q)$ in the limit of $y_1$$\rightarrow$ 0.
The term of ln$y_1$ is from the thermal spin fluctuations. 
At finite temperatures in $y_1$ $\rightarrow$ 0, the thermal fluctuations are superior to the zero-point fluctuations, that is ($t$/2)ln$y$ $\gg$ $y$ln$y$.  

We have two asymptotic solutions of Eq.~(\ref{eq1}), the logarithmic function at zero temperature limit and the Curie law at weak $y_1$ limit.
Analytically, these two asymptotic solutions cannot be smoothly connected to one another.
Thus, by solving numerically the SCR Equation~(\ref{eq1}) with the various finite values of $y_1$, we obtain $y$ against $t$ (= 0.001--0.3) for $y_1$ = 0.005--10.0 and $x_c$ = 1.0.  
And then, we analyze the numerical $y$-$t$ curves by using simpler functions. 

\section{Numerical inverse staggered spin susceptibility $y \propto \chi(Q)^{-1}$ as a function of the mode-mode coupling constant $y_1$} 

\begin{figure}[t]
 \begin{center}
 \includegraphics[width=1.0\linewidth]{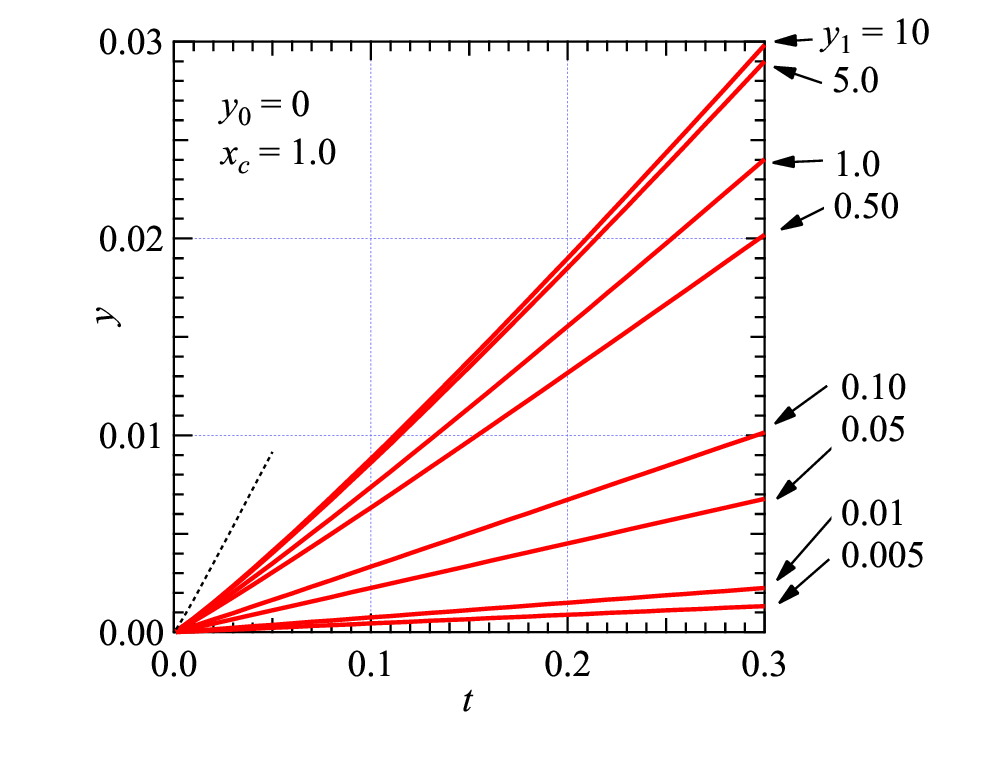}
 \end{center}
 \caption{\label{figy}
Reduced inverse staggered spin susceptibility $y$ plotted against reduced temperature $t$ = $T/T_0$ for the mode-mode coupling constant $y_1$ = 0.005--10.0. The slope of $y$ against $t$ increases with increasing $y_1$. The dotted line is the asymptotic solution $y$ = $-t$ln$|$ln$t|$/2ln$t$.  
}
 \end{figure}

Figure~\ref{figy} shows the numerical $y$ plotted against $t$ for the mode-mode coupling constant $y_1$ = 0.005--10.0. 
The dotted line is the asymptotic solution $y$ = $-t$ln$|$ln$t|$/2ln$t$, which differs from the numerical solutions at the present temperatures. 
$y$ and the slope $y/t$ increase with increasing $y_1$.
Strong mode-mode coupling suppresses the amplitude $\chi(Q) = 1/2\pi T_Ay$ of the spin fluctuations and enhances the spin fluctuation frequency $\Gamma(Q) = 2\pi T_0 y$.
The mode-mode coupling broadens the spin fluctuation spectrum toward higher frequencies. 
To understand detailed changes in the form of the $y - t$ functions with respect to $y_1$,
we analyze the numerical $y - t$ by using a power law and a Curie-Weiss law.  

\begin{figure*}[t]
 \begin{center}
\hspace*{00mm}\includegraphics[width=0.75\linewidth]{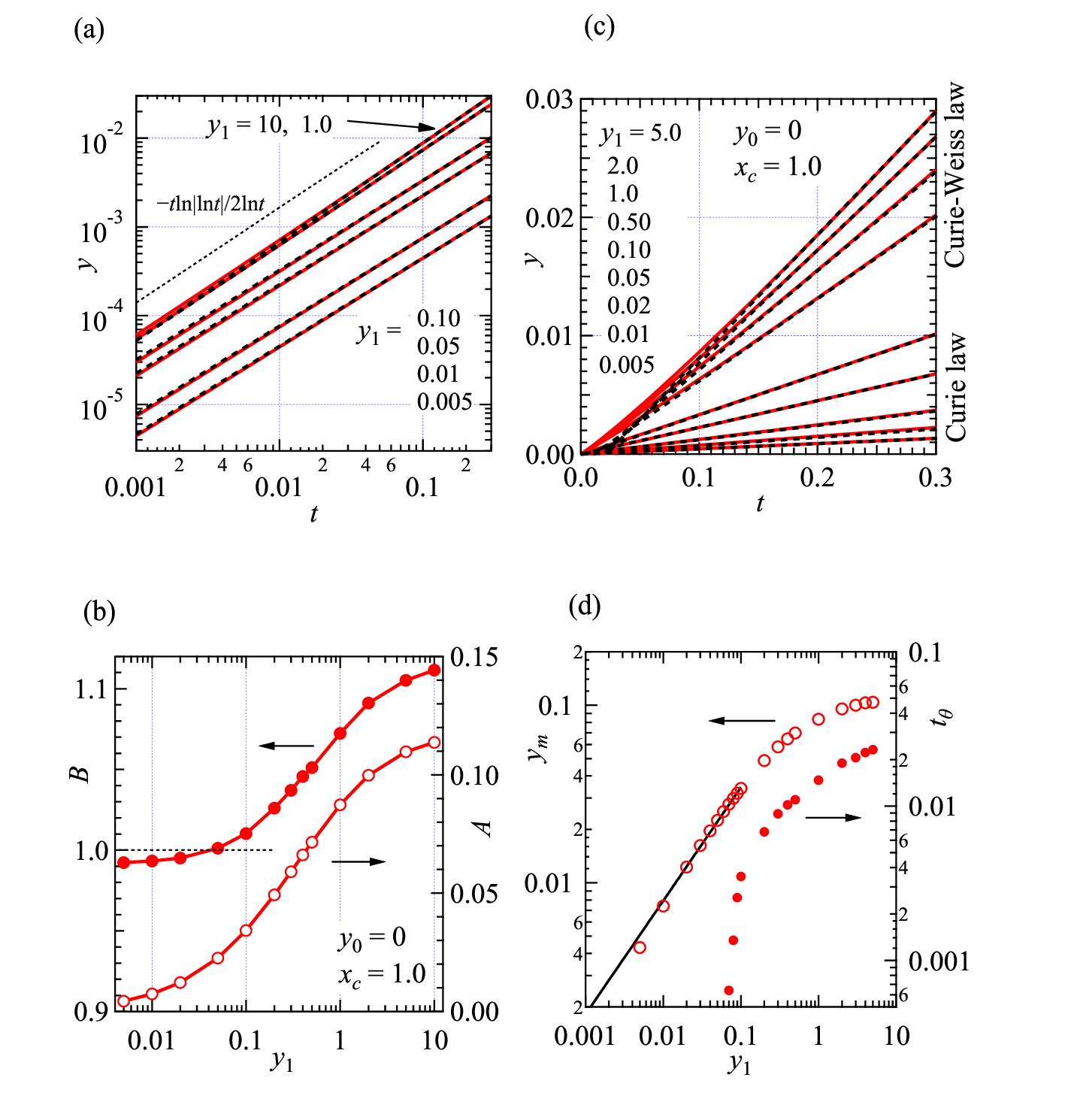}
 \end{center}
 \caption{\label{figPower}
(a) Log-log plots of numerical $y$ vs $t$ (red solid lines). The broken lines are the least-squares fitting results using the power law function $y$ = $At^{B}$ with fit parameters of $A$ and $B$.
The dotted line is the asymptotic solution $y$ = $-t$ln$|$ln$t|$/2ln$t$.
(b) Semi-logarithmic plots of the fitting results of the numerical coefficient $A$ (open circles) and the exponent $B$ (closed circles) against $y_1$.  
For $y_1 <$ 0.1, $B$ is approximately 1. As $y_1$ exceeds 0.1, $B$ increases beyond 1. 
(c) The broken lines are the least-squares fitting results using the linear functions of $y$ = $y_m(t-t_{\theta})$ with fit parameters of $y_m$ and $t_{\theta}$. 
(d) Log-log plots of the fitting results of $y_m$ (open circles) and $t_{\theta}$ (closed circles) against $y_1$.  
For $y_1 <$ 0.1, $t_{\theta}$ is nearly zero, while for $y_1 >$ 0.1 the finite $t_{\theta}$ increases with $y_1$. 
The solid line is the least-squares fitting result of $y_m = 0.15(y_1)^{0.64}$ for $y_1$ = 0.005--0.1.
}
 \end{figure*}

Figure~\ref{figPower} (a) shows the log-log plots of the numerical $y$ vs $t$ (red solid lines) from Fig.~\ref{figy}. 
All $y$-$t$ numerical curves exhibit power-law behaviors in the temperature range of $t$ = 0.001 to 0.3.
The broken lines are the least-squares fitting results using a power law function
\begin{equation}
y = At^{B}
\label{XQ} 
\end{equation}
with fit parameters of $A$ and $B$. 
The fitting results are satisfactory. 
At $t$ = 0.001--0.30, the difference between the numerical $y$ (red solid lines) and the asymptotic solution $y$ = $-t$ln$|$ln$t|$/2ln$t$ (dotted line) is large.
No crossover behavior to the asymptotic solution is found. 

Figure~\ref{figPower} (b) shows the semi-logarithmic plots of the fitting results of the numerical coefficient $A$ and the exponent $B$ against $y_1$.  
If $y_1$ is less than 0.1, $B$ is approximately 1. 
If $y_1$ exceeds 0.1, $B$ rapidly increases beyond 1.
We find a criterion $y_1$ = 0.1, which classifies the temperature dependences of $y \propto \chi(Q)^{-1}$.  
The exponent $B$ = 1.0 indicates the Curie law of $\chi(Q)$.
Thus, the $y_1$ = 0.1 serves as an indicator to distinguish whether $\chi(Q)$ follows the Curie law or not.
In Fig.~\ref{figPower} (c), the broken lines are the least-squares fitting results using a linear function
\begin{equation}
y = y_m(t-t_{\theta})
\label{XQ} 
\end{equation}
with fit parameters of $y_m$ and $t_{\theta}$. 
$y_m$ is an inverse Curie constant and $t_{\theta}$ is a Curie-Weiss temperature.
The fitting range is $t$ = 0.15--0.30. 
As the temperature $t$ decreases, the numerical $y$ with $y_1 >$ 0.1 deviates from the linear function below $t <$ 0.15,
and then $y$ goes to zero at $t$ = 0.
Thus, the finite $t_{\theta}$ does not indicate the finite temperature long range antiferromagnetic ordering.
For $y_1 <$ 0.1 in Fig.~\ref{figPower} (c) as well as Fig.~\ref{figPower} (a), 
the broken lines (Curie law) well reproduce the numerical $y$-$t$ (red solid lines) at $t$ = 0.001--0.30.

Figure~\ref{figPower} (d) shows the log-log plots of the fitting results of the inverse Curie constant $y_m$ and the Curie-Weiss temperature $t_{\theta}$ against $y_1$.
If $y_1$ is less than 0.1, $t_{\theta}$ is approximately 0 and then $y_m$ is of a power law as a function of $y_1$.
If $y_1$ exceeds 0.1, $t_{\theta}$ increases with increasing $y_1$.
Thus, the criterion value of $y_1$ = 0.1 is the same in both power-law analysis and Curie-Weiss analysis.

In Fig.~\ref{figPower} (d), the solid line is the least-squares fitting result using $y_m = {\alpha}(y_1)^{\beta}$ for $y_1 <$ 0.1.
We obtain ${\alpha}$ = 0.15 and ${\beta}$ = 0.64, and then $y = 0.15(y_1)^{0.64}t$ for $y_1$ = 0.005--0.1. 
The fitting result is satisfactory because of the low chi-square value $\chi^2$ = 1.7$\times$10$^{-6}$. 
The Curie law of 
\begin{equation}
\chi(Q) = \frac{1}{{\alpha}(y_1)^{\beta}}\frac{T_0}{2T_A}\frac{1}{T}
\label{eqQCP} 
\end{equation}
holds in the weak mode-mode coupling regime ($y_1 <$ 0.1).  

We also find that the analytical function $y/t = 0.1755(y_1)^{0.565}$ reproduces the Moriya-Takahashi-Ueda SCR result on $y/t$ for $y_1$ = 0--5.5 in the absence of the zero-point fluctuations.
Figure 3 demonstrates $y/t = 0.1755(y_1)^{0.565}$, which reproduces the numerical data in Fig. 2 in Ref.~\cite{MTU}.
\begin{figure}[t]
 \begin{center}
\hspace*{00 mm}\includegraphics[width=0.70\linewidth]{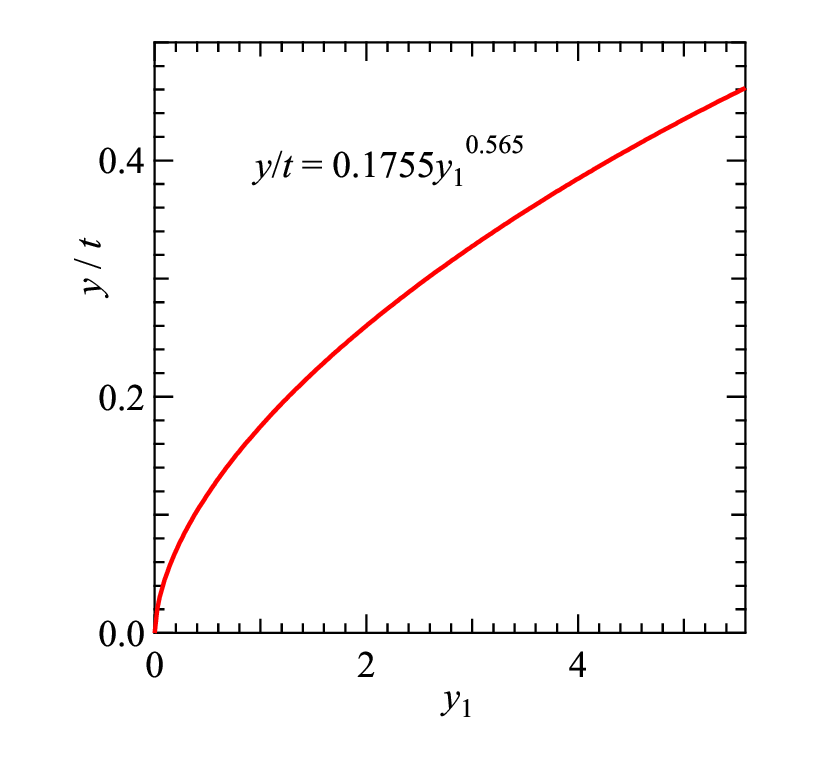}
 \end{center}
 \caption{\label{fig3}
The solid curve demonstrates $y/t = 0.1755(y_1)^{0.565}$ for the numerical SCR solution in the absence of zero-point fluctuations (see Figure 2 in Ref.~\cite{MTU}).
}
 \end{figure} 

For $y_1 >$ 0.1, the Curie-Weiss law at higher temperatures ($t >$ 0.15) is 
\begin{equation}
\chi(Q) = \frac{C}{T - {\Theta}},
\label{eqCW} 
\end{equation}
where $C$ = $T_0/2y_mT_A$ and $\Theta$ = $t_{\theta}T_0$. 
The values of $y_m$ and $t_{\theta}$ as functions of $y_1 (> 0.1)$ are given in Fig.~\ref{figPower} (d).
When $y_1$ exceeds beyond 1.0, the rate of increase for $y_m$ and $t_{\theta}$ slows down. 

Let us examine how much the zero-point fluctuations contribute in the SCR Eq.~(\ref{eq1}) for a few typical values of $y_1$
and the origin of the Curie-Weiss temperature.   
The SCR Equation~(\ref{eq1}) is expressed by $y$ = $\frac{y_1}{2}[ Z(t) + H(t)]$, where
\begin{equation}
Z(t) = y(t){\rm ln}y(t) - cy(t),
\label{eqZ}
\end{equation}
and
\begin{equation}
H(t) = \frac{t}{2}\Big\{{\rm ln}\frac{x_c^2 + y(t)}{y(t)}-{\rm ln}\frac{x_c^2 + y(t) + t/6}{y(t) + t/6}\Big\}. 
\label{eqH}
\end{equation}
$H(t)$ shows the contribution from the thermal fluctuations, while $Z(t)$ shows the counter contribution from the zero-point fluctuations.
The complete cancelation of the zero-point fluctuation $Z(t)$ and the thermal fluctuation $H(t)$ yields the asymptotic logarithmic $t$ dependence of $y(t)$ in the zero temperature limit~\cite{WM,IM1}. 
In Equation~(\ref{eq1}), $Z(t)$ and $H(t)$ are in competition with each other.      
 \begin{figure}[t]
 \begin{center}
 \includegraphics[width=1.0\linewidth]{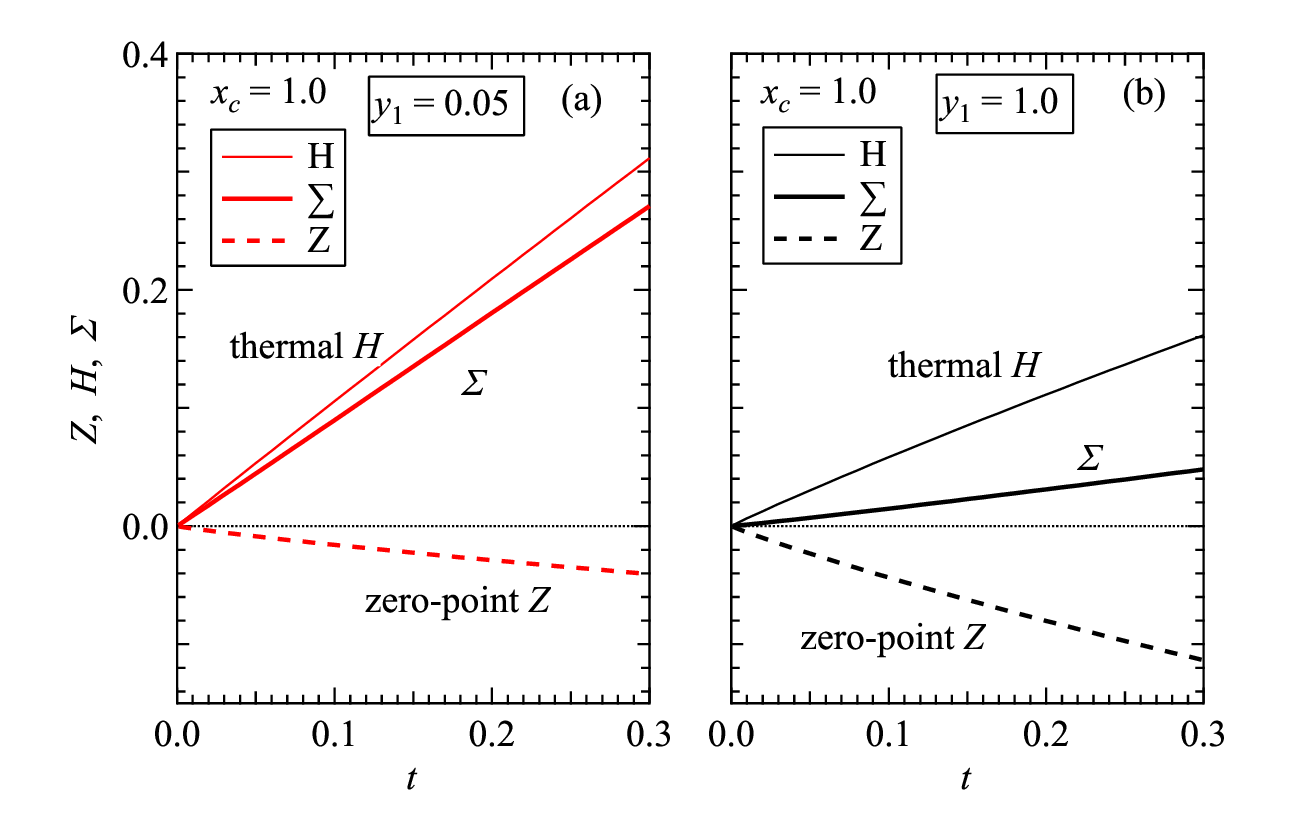}
 \end{center}
 \caption{\label{fig4}
Zero-point quantum fluctuations $Z(t)$, thermal fluctuations $H(t)$, and the summation $\Sigma$ = $Z(t)$ + $H(t)$ plotted against $t$ for $y_1$ = 0.05 (a) and 1.0 (b).
The dotted horizontal lines indicate zero on the vertical axis.
}
 \end{figure}  

The numerical solutions $y(t)$ ($y_1$ = 0.05 and 1.0) in Fi. 1 are substituted for those in Eq.~(\ref{eqZ}) and Eq.~(\ref{eqH})
to derive the individual contributions in Eq.~(\ref{eq1}).
Figure~\ref{fig4} shows $Z(t)$, $H(t)$, and the summation of $\Sigma$ = $Z(t)$ + $H(t)$ against $t$ for $y_1$ = 0.05 (a) and 1.0 (b).
For a small $y_1$ = 0.05, $H(t)$ is larger than $|Z(t)|$, whereas
for a large $y_1$ = 1.0, $H(t)$ is comparable to $|Z(t)|$. 
The incomplete cancellation of $Z(t)$ and $H(t)$ results in the Curie-Weiss law with the finite $t_{\theta}$. 
In the strong mode-mode coupling regime with $y_1 >$ 0.1, the zero-point fluctuation $Z(t)$ at high temperatures plays a significant role
and leads to the finite Curie-Weiss temperature. 
\\
\section{Nuclear spin-lattice relaxation rate due to two-dimensional antiferromagnetic spin fluctuations} 
For experimental application, let us consider the nuclear spin-lattice relaxation rate 1/$T_1$ due to the two-dimensional antiferromagnetic spin fluctuations.
For the two-dimensional nearly antiferromagnets, 1/$T_1$ is expressed as 
\begin{equation}
\frac{1}{T_1} = \frac{h}{k_B}\Big(\frac{\gamma_N}{2\pi}\Big)^2\frac{A_{hf}^2}{T_0T_A}\frac{T}{y}
\label{T1}
\end{equation}
where $\gamma_N$ is the nuclear gyromagnetic ratio and $A_{hf}$ is the hyperfine coupling constant~\cite{MTU,IM1}. 
We can estimate $y$ from the experimental $T_1T$ as
\begin{equation}
y = (T_1T)S,
\label{T1T}
\end{equation}
where $S$ = $(h/k_B)(\gamma_N/2\pi)^2A_{hf}^2/T_0T_A$. 

\begin{figure}[t]
 \begin{center}
  \includegraphics[width=1.0\linewidth]{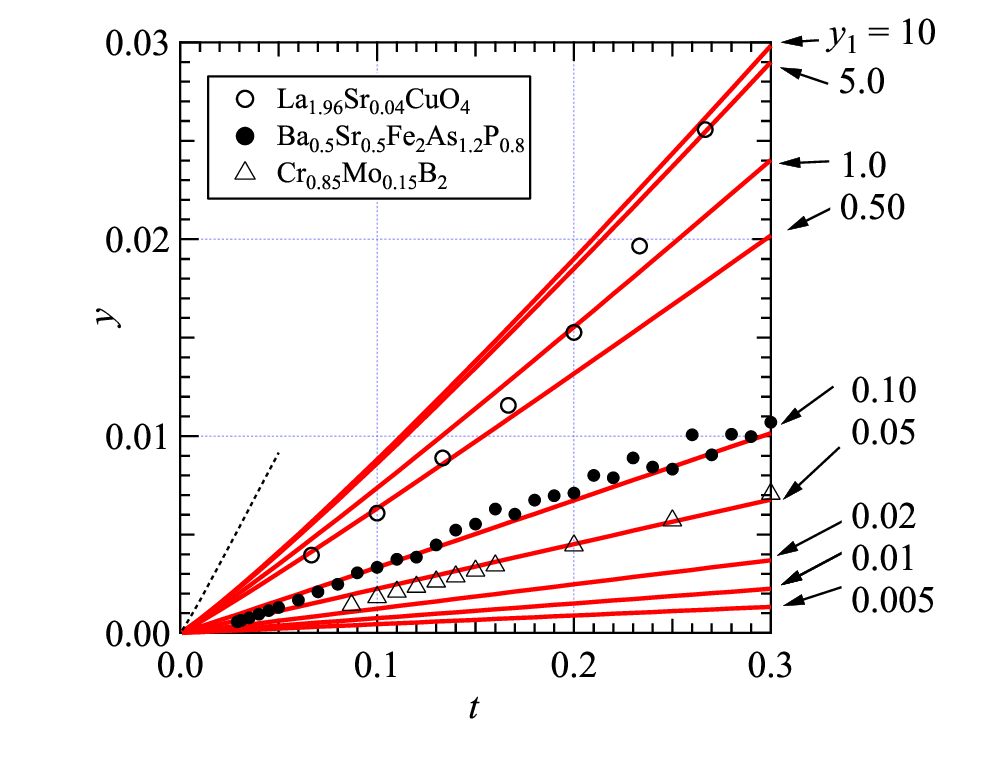}
 \end{center}
\caption{\label{fig5}
Experimental $y_{ex} \equiv (T_1T)S$ plotted against $t = T/T_0$, which are estimated from $^{63}$Cu nuclear spin-lattice relaxation rate 1/$^{63}T_1$ for lightly-doped La$_{1.96}$Sr$_{0.04}$CuO$_{4}$ (open circles)~\cite{Imai}, $^{31}$P nuclear spin-lattice relaxation rate 1/$^{31}T_1$ for optimally-doped Ba$_{0.5}$Sr$_{0.5}$Fe$_2$As$_{1.2}$P$_{0.8}$ (closed circles)~\cite{Itoh3},
and $^{11}$B nuclear spin-lattice relaxation rate 1/$^{11}T_1$ for non-superconducting Cr$_{0.85}$Mo$_{0.15}$B$_2$ (open triangles)~\cite{Yoshimura}.
The coefficients are $S$ = 0.27 s$^{-1}$K$^{-1}$ for LSCO, 0.33$\times$10$^{-3}$ s$^{-1}$K$^{-1}$ for Ba122, and 
0.40$\times$10$^{-3}$ s$^{-1}$K$^{-1}$ for CrBMo.
The~dotted line is the asymptotic solution $y$ = $-t$ln$|$ln$t|$/2ln$t$.
\label{T1TS}
}
 \end{figure}

We analyze $^{63}$Cu nuclear spin-lattice relaxation time $^{63}T_1$ for lightly-doped La$_{1.96}$Sr$_{0.04}$CuO$_{4}$ (LSCO)~\cite{Imai}, 
$^{31}$P nuclear spin-lattice relaxation time $^{31}T_1$ for optimally-doped Ba$_{0.5}$Sr$_{0.5}$Fe$_2$As$_{1.2}$P$_{0.8}$ (Ba122)~\cite{Itoh3}, and
$^{11}$B nuclear spin-lattice relaxation time $^{11}T_1$ for non-superconducting Cr$_{0.85}$Mo$_{0.15}$B$_2$ (CrBMo)~\cite{Yoshimura}. 
Figure~\ref{T1TS} shows the experimental $y_{ex} \equiv (T_1T)S$ against $t = T/T_0$ for LSCO (open circles), Ba122 (closed circles), and CrBMo (open triangles). 
The reduced temperature $t$ = $T/T_0$ is defined by using $T_0\sim$ 1500 K for LSCO, 1000 K for Ba122, and 1000 K for CrBMo. 
We adopt $T_A \sim$ 1500 K for LSCO, 4200 K for Ba122, and 650 K for CrBMo, tentatively.
The values of $S$ = 0.27 s$^{-1}$K$^{-1}$ for LSCO, 0.33$\times$10$^{-3}$ s$^{-1}$K$^{-1}$ for Ba122, and 
0.40$\times$10$^{-3}$ s$^{-1}$K$^{-1}$ for CrBMo are consistent with estimation based on hyperfine coupling constants. 

The theoretical $y$-$t$ curves with $y_1$ = 0.5$\sim$10.0 cover the experimental $y_{ex}$ for LSCO.
The experimental $y_{ex}$ data in such a wide range of the $y_1$ values may be due to the frequency distribution of $^{63}T_1$~\cite{Imai3}.
LSCO can be classified as a system with strong mode-mode coupling.
The theoretical $y$-$t$ curves with $y_1$ = 0.05$\sim$0.1 cover the experimental $y_{ex}$ for Ba122.
The SCR analysis in the absence of the zero-point fluctuations has been performed for BaFe$_2$(As$_{1-x}$P$_{x}$)$_2$~\cite{Ishida}.
Ba122 can be classified as a system with weak or intermediate mode-mode coupling.
The theoretical curve with $y_1\sim$ 0.05 reproduces the experimental $y_{ex}$ for CrBMo. 
CrBMo can be classified as a system with weak mode-mode coupling.
Thus, these compounds can be classified based on the strength of $y_1$ in the vicinity of the two-dimensional antiferromagnetic QCP.
\\
\section{Conclusion}
In conclusion, the effect of the zero-point quantum fluctuations on the temperature dependence of the staggered spin susceptibility $\chi(Q)$ depends on the strength of the mode-mode coupling constant $y_1$ at the two-dimensional antiferromagnetic QCP.   
In the weakly coupling regime ($y_1 <$ 0.1), the predominant thermal spin fluctuations lead to a Curie law of $\chi(Q)$.  
In the strongly coupling regime ($y_1 >$ 0.1), the zero-point spin fluctuations have an influence comparable to that of the thermal fluctuations in determining $\chi(Q)$.
As to experimental studies, the absence of the Curie-Weiss temperature ($t_{\theta}$ = 0) can serve as identification of the QCP ($y_0$ = 0) for the weak mode-mode coupling systems ($y_1 <$ 0.1). 




\end{document}